\documentclass[preprint,prl,aps,amsmath,amssymb,color,superscriptaddress]{revtex4}
\usepackage{stmaryrd}
\usepackage{amsmath}
\usepackage{amssymb}
\usepackage{graphicx}
\usepackage{dcolumn}
\usepackage{bm}
\usepackage{tabularx}
\usepackage{color}
\usepackage{booktabs}
\usepackage{multirow}
\begin{document}
	
	\begin{titlepage}
		\title{Magnetoelectric effect in multiferroic metals via a direct spin-charge interaction}
		
		\author{Zefei Han}
		\affiliation{Key Lab of advanced optoelectronic quantum architecture and measurement (MOE), and School of Interdisciplinary Science, Beijing Institute of Technology, Beijing 100081, China}
		
		\author{Haojin Wang}
		\affiliation {Key Lab of advanced optoelectronic quantum architecture and measurement (MOE), and School of Interdisciplinary Science, Beijing Institute of Technology, Beijing 100081, China}
		
		\author{Yuanchang Li}
		\email{yuancli@bit.edu.cn}
		\affiliation{Key Lab of advanced optoelectronic quantum architecture and measurement (MOE), and School of Interdisciplinary Science, Beijing Institute of Technology, Beijing 100081, China}
		
		\date{\today}
		
		\begin{abstract}
Much is known about the magnetoelectric effect of multiferroic insulators, yet little is understood about multiferroic metals. In this work, we propose a stacking engineering strategy based on monolayer magnets to construct multiferroic metals, with validation through first-principles calculations on six experimentally synthesized materials. Such multiferroic metals exhibit predominantly linear magnetoelectric response, originating from direct spin-charge interactions as a result of external field-modulated Fermi energy. This fundamentally differs from spin-charge-lattice or spin-orbit coupling mechanisms in multiferroic insulators, offering application advantages in the field of high-speed response. We derive a universal formula for understanding the magnetoelectric coupling in these multiferroic metals. Our work provides insights for exploring magnetoelectric coupling mechanisms and designing functional materials with strong magnetoelectric coupling.
		\end{abstract}

		\maketitle
		\draft
		\vspace{2mm}
	\end{titlepage}
	
Multiferroics, which combine ferroelectricity and magnetism, constitute a fascinating class of magnetoelectric systems\cite{multiferro-1}. Their capacity to modulate magnetic properties via electric fields, and conversely, to modulate electrical properties via magnetic fields, confers a wide range of potential applications in fields such as spintronic devices, memories, and data storage\cite{multiferro-2,multiferro-3}. Traditionally, research on magnetoelectric multiferroics has primarily focused on insulating materials with well-defined bandgaps, paying little attention to metals. This disregard stems from two major considerations. On the one hand, the screening of free charge carriers in metals repels applied electric fields to their surfaces, hindering the observation of significant electric field effects. Yet when the characteristic dimension of the metal is reduced to a scale comparable to the electric field penetration depth, electric field effects can become pronounced. With the realization of room-temperature electric field modulation of coercivity in FePt and FePd films\cite{Weisheit}, much progress has been made in the electrically controlled magnetism of metals\cite{Matsukura,DuanCG}. On the other hand, metallicity and ferroelectricity are difficult to reconcile. Although the ferroelectric metal was proposed sixty years ago\cite{Anderson}, it was only recently that LiOsO$_3$\cite{ShiY} was confirmed as the first ferroelectric-like metal, followed by ferroelectric switching in conductive WTe$_2$ flakes\cite{wte2-exp}.

Van der Waals (vdW) multilayers offer new avenues for exploring metallic magnetoelectric multiferroics. Their uniqueness lies in the presence of sliding ferroelectricity$-$a spontaneous out-of-plane polarization arising from lateral interlayer sliding rather than ionic displacement\cite{slide-wmh}. As such ferroelectricity is a consequence of stacking, it is decoupled from in-plane nature, thereby enabling coexistence with in-plane metallicity\cite{efield-cal,ZhouWX}. This naturally paves the way for realizing multiferroic metals that combine metallicity, ferroelectricity, and magnetism through stacking engineering of monolayer magnets. Encouragingly, since the experimental confirmation of intrinsic magnetism in layered CrGeTe$_3$\cite{Cr2Ge2Te6} and CrI$_3$\cite{CrI3}, a number of two-dimensional metallic magnets have been  reported, such as NbTe$_2$\cite{mono1-nbte2,mono2-nbte2}, Fe$_3$GeTe$_2$\cite{Deng,Ghosh}, VSe$_2$\cite{Bonilla,Wines}, MnSe$_2$\cite{Hara,LiJW}, CrS$_2$\cite{Xiao,WangC} and CrTe$_2$\cite{ZhangX}.

Despite progress, research on multiferroic metals remains nascent compared to multiferroic insulators\cite{FEM-1,FEM-2,Os2Se3,AA',YuPu,Brehin}. Multiferroic metals themselves are scarce, let alone their response to external electric/magnetic fields. It remains unknown what kind of magnetoelectric behaviour they exhibit, or whether they can be utilized in magnetoelectric devices like multiferroic insulators\cite{response}. Such insights hold significant scientific value for deepening our comprehension of the interplay between multiple orders within multiferroics, while also offering practical implications for the development of next-generation magnetoelectric devices.

In this work, we first take 1$T$-NbTe$_{2}$ as a prototype and reveal through first-principles calculations that its antiparallel stacked bilayer is a sliding multiferroic metal. Lateral sliding induces interlayer charge transfer, which on the one hand generates an out-of-plane polarization $P$, and on the other hand prevents the complete compensation of antiparallel magnetic moments, giving rise to a non-zero total moment $M$. We have subsequently investigated its magnetoelectric properties by tracking the evolution of $M$ and $P$ under applied electric and magnetic fields. The results show predominantly linear magnetoelectric behavior with a coupling parameter of $\sim$1.4 ps/m. It originates from direct spin-charge interactions arising when external electric/magnetic fields modulate the Fermi level, which fundamentally differs from insulating multiferroics typically relying on lattice-mediation or spin-orbit coupling\cite{XuML,DongS}. Finally, we derive a universal formula describing the magnetoelectric coupling in such multiferroic metals. We also validate the universal applicability of the construction strategy and magnetoelectric mechanism across material systems with different architectures and magnetic configurations.
	
Density functional theory (DFT) calculations were performed using the Vienna Ab initio Simulation Package (VASP)\cite{vasp2} with the Perdew-Burke-Ernzerhof (PBE) exchange-correlation functional\cite{PBE}. A Hubbard $U$ term of 2.91 eV was applied for Nb 4$d$ orbitals, as previously employed\cite{nbte2um}. The vdW interactions were considered using the Grimme's DFT-D3 (BJ) scheme\cite{bj}. Electron-ion interaction was described by the projector augmented wave method\cite{PAW,PAW2} with an energy cutoff of 500 eV. A vacuum layer at least 20 \AA\ was added to avoid spurious interactions between two neighboring images. A 17 $\times$ 17 $\times$ 1 $\Gamma$-centered $k$-mesh was used to sample the Brillouin zone with a second-order Methfessel-Paxton smearing of 0.05 eV (see Supplemental Materials\cite{SI} for convergence studies). The convergence criteria of energy and force were set to 10$^{-6}$ eV and 10$^{-3}$ eV/\AA, respectively. The phonon spectrum was calculated using a 4 $\times$ 4 $\times$ 1 supercell within the density functional perturbation theory as implemented in the PHONOPY package\cite{phonon}. Polarization was computed using the dipole method\cite{dipole-yyg}. The out-of-plane electric and magnetic fields were implemented using the VASP EFILED and BEXT tags, with dipole corrections employed to evaluate the polarizations.

	\begin{figure*}[htb]
		\includegraphics[width=0.98\columnwidth]{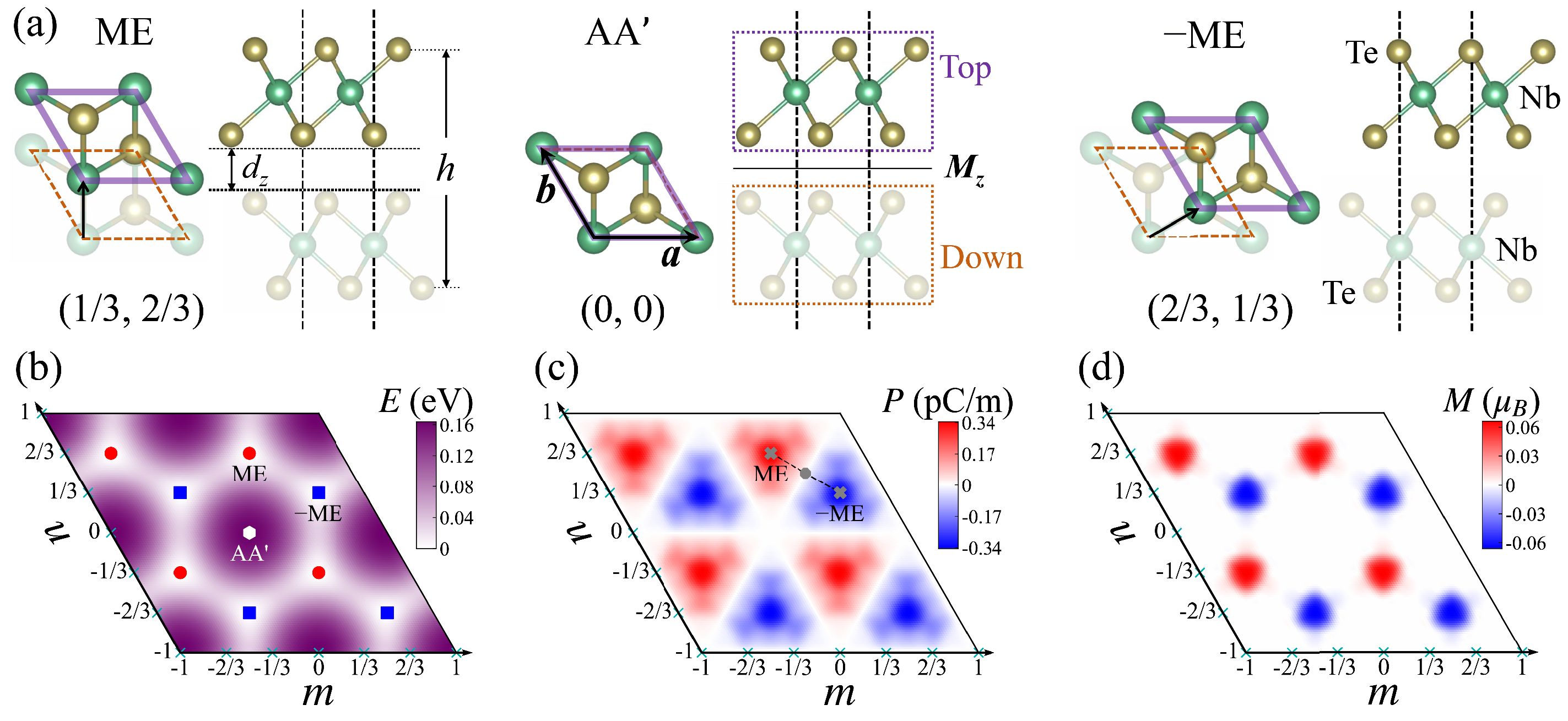}
		\caption{\label{fig:fig1} (a) Geometric structures for the ME state (Left), antiparallel AA' stacking (Middle), and $-$ME state of 1$T$-NbTe$_2$ bilayer. In the left and right panels, the black arrows connecting the orange and purple rhombuses denote the interlayer sliding vector \textit{\textbf{l}} = $m$\textit{\textbf{a}} + $n$\textit{\textbf{b}}, with ($m$, $n$) = $(\frac{1}{3},\frac{2}{3})$ and $(\frac{2}{3},\frac{1}{3})$, respectively. $h$ and $d_z$ represent the bilayer thickness and the interlayer spacing, respectively. In the middle panel, $M_z$ denotes mirror symmetry between the top and down layers. Green and yellow balls denote Nb and Te atoms, respectively. (b) Total energy, (c) ferroelectric polarization, and (d) total magnetic moment as a function of ($m$, $n$) for the sliding 1$T$-NbTe$_2$ bilayer. In (b), the four equivalent red (blue) positions are labeled ME ($-$ME). In (c), the black dashed line indicates the ferroelectric switching path and the grey dot marks the transition state position.}
	\end{figure*}
	
We begin with $1$T$-$NbTe$_2$ to illustrate the strategy for constructing multiferroic metals and the corresponding magnetoelectric properties. Its monolayer features a trigonal phase with space group $P\overline{3}m1$. Our calculations indicate that it is a ferromagnetic metal with a total moment of 0.57 $\mu_B$, in line with the previous study\cite{nbte2um}. Regarding the bilayer, we first consider the antiparallel AA' stacking configuration depicted in the middle panel of Fig. 1(a), which exhibits the mirror symmetry about the $M_z$ plane. However, the appearance of soft phonon modes (see Fig. S2\cite{SI}) indicates structural instability. This is because the stacking forces the large Te atoms of the top and down layers to directly face each other, thereby enhancing steric repulsion. To mitigate such repulsion, lateral interlayer sliding occurs, which can be characterized by the vector \textit{\textbf{l}} = $m$\textit{\textbf{a}} + $n$\textit{\textbf{b}}, where \textit{\textbf{a}} and \textit{\textbf{b}} are in-plane lattice vectors as illustrated in Fig. 1(a). Under periodic boundary conditions, $m$ and $n$ range from ($-$1, 1). For a given \textit{\textbf{l}}, the equilibrium interlayer spacing is determined through structural optimization. As such, a pair ($m$, $n$) uniquely defines a sliding structure. The left and right panels of Fig. 1(a) present the optimized configurations for $(\frac{1}{3},\frac{2}{3})$ and $(\frac{2}{3},\frac{1}{3})$ bilayer, respectively.

Figures 1(b)$-$1(d) summarize the total energy, ferroelectric polarization, and total magnetic moment of the bilayer as a function of ($m$, $n$). The total energy exhibits eight minima, namely, $(\frac{1}{3},\frac{2}{3})$, $(\frac{1}{3},-\frac{1}{3})$, $(-\frac{2}{3},\frac{2}{3})$, $(-\frac{2}{3},-\frac{1}{3})$, and $(\frac{2}{3},\frac{1}{3})$, $(\frac{2}{3},-\frac{2}{3})$, $(-\frac{1}{3},\frac{1}{3})$, $(-\frac{1}{3},-\frac{2}{3})$. They all possess spontaneous polarization of 0.34 pC/m and total magnetic moment of 0.066 $\mu_B$, but with the former four (the red dots in Fig. 1(b), hereafter referred to as the ME state) along the +$z$ direction and the latter four (the blue squares in Fig. 1(b), hereafter referred to as the $-$ME state) along the $-$$z$ direction. Such polarization is on a par with that of the WTe$_{2}$ bilayer (0.35 pC/m)\cite{wte2-wmh} but exceeds that of the CrI$_{3}$ bilayer (0.18 pC/m)\cite{cri3}. Employing the CI-NEB method\cite{cineb}, we have identified a ferroelectric switching path depicted by the black dashed line in Fig. 1(c), which gives a barrier of 15.5 meV per formula. This value is comparable to typical sliding ferroelectrics, such as 15 meV for MoS$_{2}$ bilayer\cite{mos2} and 12 meV for SiC bilayer\cite{sic}. As switching does not involve bond breaking, the barrier of sliding ferroelectrics is an order of magnitude lower than that of displacive ferroelectrics such as 170 meV and 475 meV per formula, respectively, for BaTiO$_3$ and BiFeO$_3$\cite{barrier}.

	\begin{figure}[htb]
		\includegraphics[width=0.50\columnwidth]{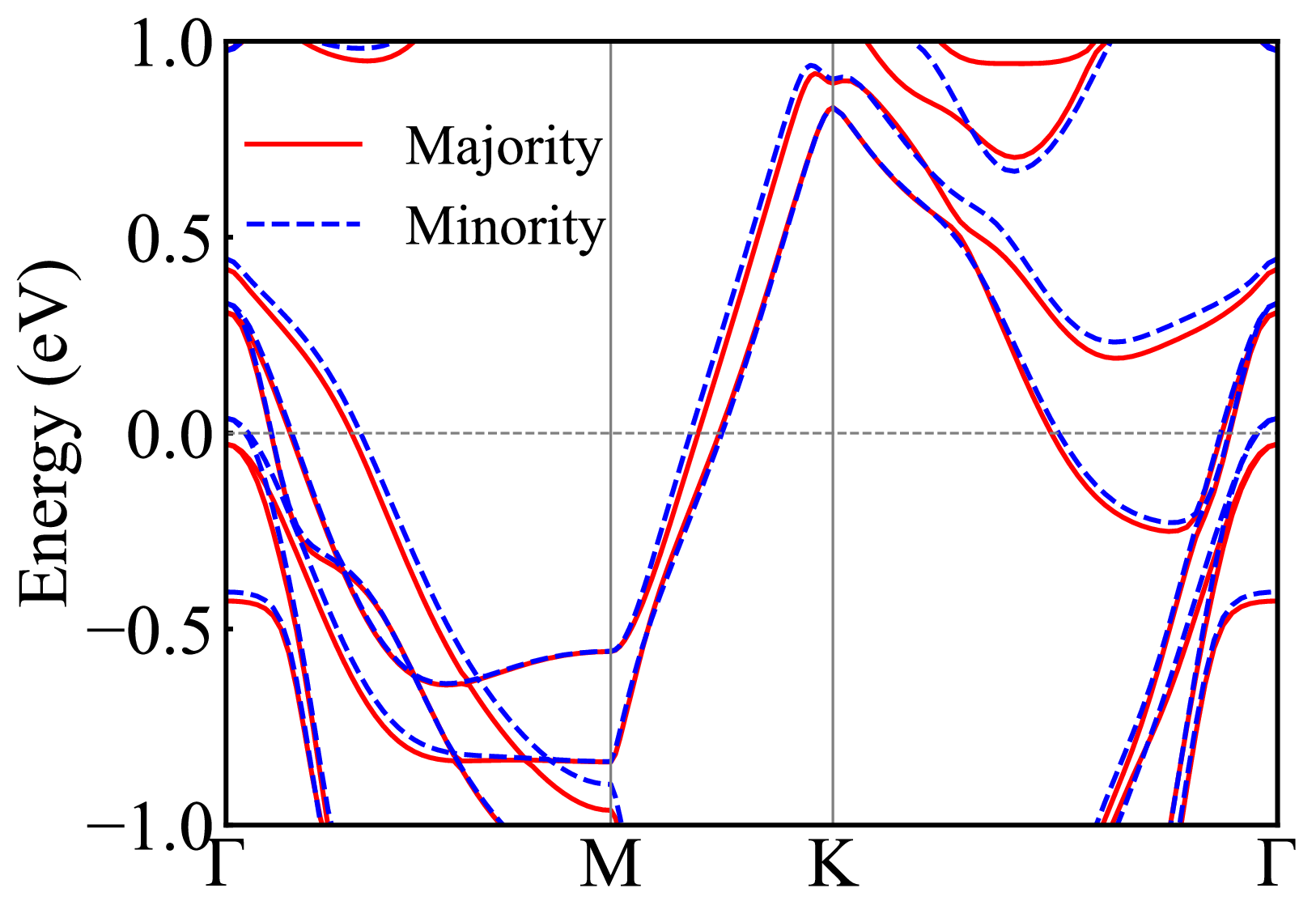}
		\caption{\label{fig:fig2} Spin-resolved band structure for the ME state, with spin-majority in red solid lines and spin-minority in blue dashed lines, respectively. The Fermi level is set at energy zero.}
	\end{figure}
	
Figure 2 presents the band structure of the ME state. It is a metal, as further confirmed by the hybrid functional (see Fig. S3\cite{SI}). We find an intriguing spin-layer locking phenomenon, wherein the spin-majority is fully contributed by the down layer and the spin-minority is fully contributed by the top layer. This causes antiparallel coupling of magnetic moments between the top and down layers, and the magnitude of the net magnetic moment is determined by the incomplete interlayer compensation. The more incomplete the interlayer compensation, the more pronounced the band spin-splitting, corresponding to a larger net moment. For the high-symmetry AA' configuration, compensation is complete, resulting in a zero total moment. Moreover, the ME state possesses an out-of-plane easy axis with a magnetic anisotropy energy of 123 $\mu$eV relative to the in-plane axis.
	
Putting all these results together, the ground-state of the sliding 1$T$$-$NbTe$_2$ bilayer combines ferroelectricity, magnetism and metallicity. It is therefore a rare multiferroic metal.

	\begin{figure}[htb]
		\includegraphics[width=0.8\columnwidth]{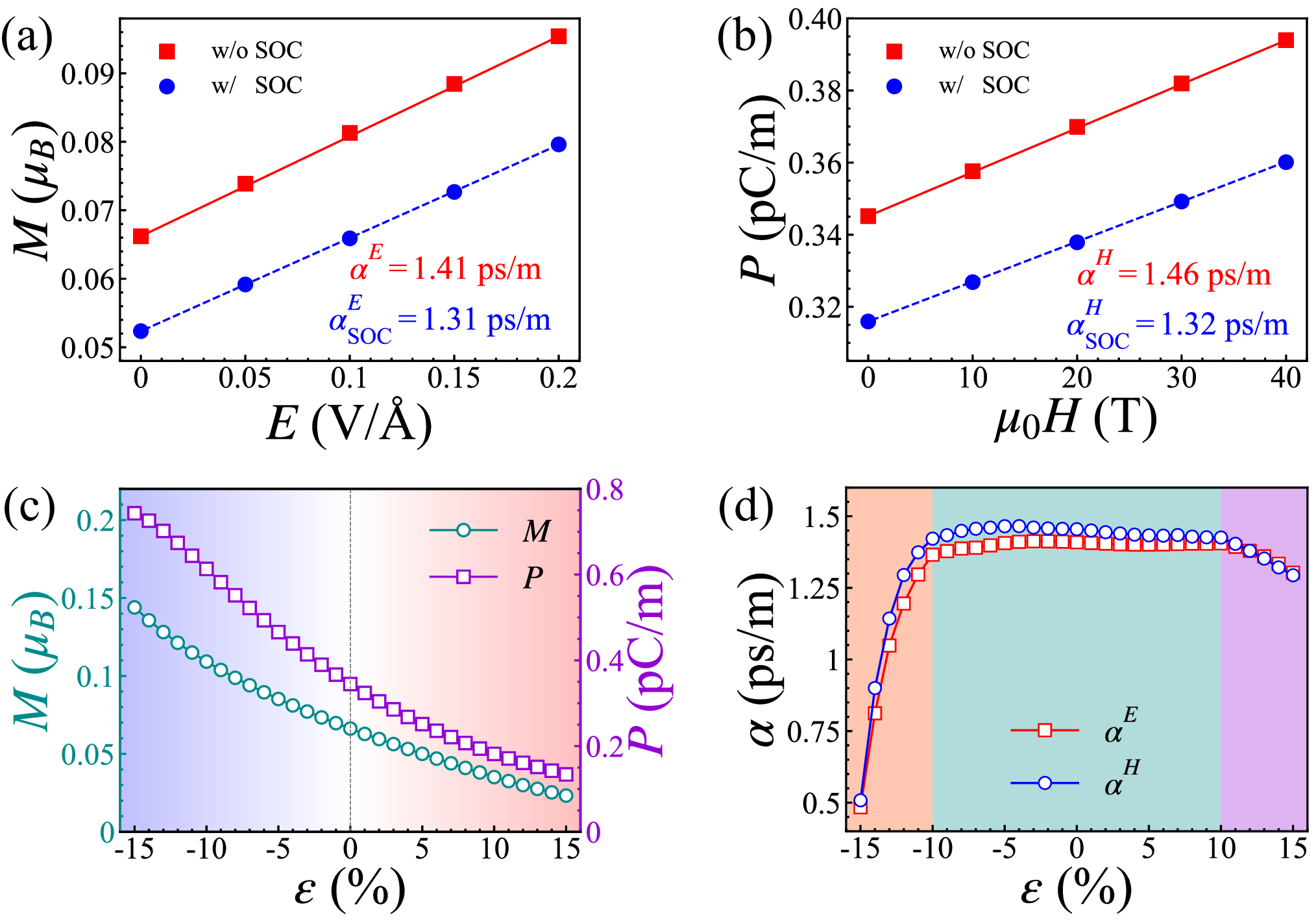}
		\caption{\label{fig:fig3} Variation of (a) total magnetic moment ($M$) with applied electric field ($E$), and (b) polarization ($P$) with applied magnetic field ($\mu_0 H$), excluding and including spin-orbit coupling (w/o and w/ SOC). Data points are calculated from first-principles calculations, with curves representing linear fits. Strain-dependent (c) total magnetic moment and polarization, and (d) magnetoelectric coupling parameters $\alpha^{E}$ and $\alpha^{H}$.}
	\end{figure}
	
Next, we investigate the magnetoelectric behavior by tracking the evolution of $M$/$P$ under electric/magnetic fields. As in-plane electric/magnetic fields are ineffective for $M$/$P$, the magnetoelectric coupling tensor here degrades to a constant $\alpha^E$/$\alpha^H$, unlike in multiferroic insulators\cite{Wang-arXiv}. Test calculations confirm that no change in $P$ is observed even under an in-plane magnetic field as high as 40 T.
	
	\begin{figure}[htb!]
		\includegraphics[width=0.6\columnwidth]{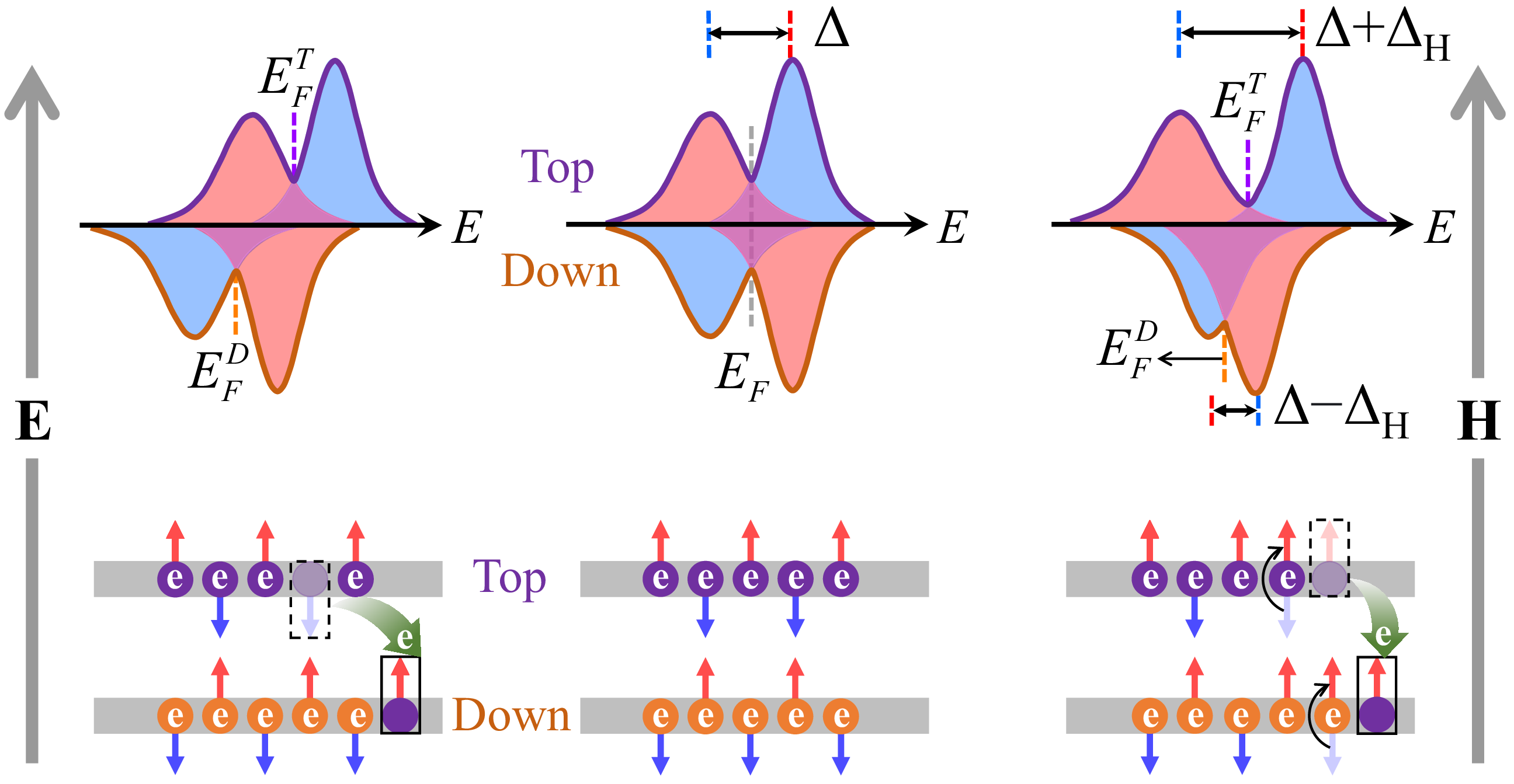}
		\caption{\label{fig:fig4} Schematic of the Fermi-energy-modulation mechanism for the magnetoelectric response of the 1$T$-NbTe$_2$ bilayer, based on first-principles calculations (see the Supplementary Materials for details\cite{SI}). Upper row: Red and blue represent spin-up and spin-down electronic states, respectively. Lower row: Red-upward and blue-downward arrows denote spin-up and spin-down electrons, respectively; green arrows indicate interlayer charge transfer. Black-dashed and black-solid boxes represent electrons donated and received. Black curved arrows indicate spin flipping within layers. Middle panel: For highly symmetric AA' bilayers, although monolayer NbTe$_2$ exhibits spin-splitting $\Delta$, it is of equal magnitude but opposite sign between the top and down layers, forming an antiferromagnet with a zero total moment. $E_F$ denotes the Fermi-level position. Left panel: An applied $z$-orientated electric field induces an interlayer potential difference, causing the electronic states of the top layer to shift upwards relative to the down layer and resulting in a difference in individual Fermi levels ($E^T_F$ vs. $E^D_F$ in the top-left). To maintain an identical Fermi level across the system, a portion of electrons transfer from the top to down layer. This process is accompanied by spin flipping, preventing complete cancellation of interlayer moments and altering the total magnetization. Right panel: An applied $z$-orientated magnetic field produces dual effect. It excites partial spins to align with the external magnetic field while introducing Zeeman splitting $\Delta_H$. As a result, the spin splitting in the top layer broadens while that in the down layer narrows. This asymmetry creates a difference in Fermi levels between layers, thereby driving interlayer charge transfer and altering the polarization.}
	\end{figure}

Figures 3(a) and 3(b) present the responses of $M$ and $P$ to electric and magnetic fields, respectively, both of which exhibit predominantly linear characteristics. Using $\mu_{0} \Delta M=\alpha^{E} E V$ ($\mu_0$ is the permeability of vacuum and $V$ is the unit cell volume) and $\Delta P=\alpha^{H} H h$ [$h$ = 10.53 \AA\ as illustrated in Fig. 1(a)], it yields $\alpha^{E}$=1.41 ps/m and $\alpha^ {H}$=1.46 ps/m. Approximately 1 ps/m is a typical value for type-I multiferroics\cite{Wang-arXiv}. When spin-orbit coupling is considered, the linear magnetoelectric response remains unchanged, with a slight decrease in the coupling parameters to $\alpha^{E}=1.31$ ps/m and $\alpha^{H}=1.32$ ps/m. Note that within our numerical accuracy limits, no induced atomic displacements or structural distortions are observed from the applied electric/magnetic fields, indicating a strictly zero lattice-mediated magnetoelectric response.

We also examine the piezoelectric property by manually altering the interlayer spacing $d_{z}$ [see Fig. 1(a) for $d_{z}$]. Here, we define the strain as $\varepsilon=\frac{d_{z}-d_{0}}{d_{0}} \times 100\%$, where $d_{0}$ is the equilibrium interlayer spacing. Figure 3(c) presents the dependence of $M$ and $P$ on $\varepsilon$. They follow very similar trends, exhibiting (sub)linear relationships with $\varepsilon$. Meanwhile, the coupling parameters exhibit insensitivity to $\varepsilon$, as demonstrated in Fig. 3(d), where $\alpha^{E}$ and $\alpha^{H}$ remain virtually unchanged within the range ($-$10\%, 10\%).

Figure 4 elucidates the magnetoelectric response mechanism of the 1$T$$-$NbTe$_2$ bilayer, which we term Fermi-energy-modulation. For the high-symmetry AA' configuration, not only is there no net charge transfer between layers, but the magnetic moments of the top and down layers also completely compensate each other. Consequently, in-plane metallicity coexists with out-of-plane antiferromagnetism. Applying a $z$-orientated electric field induces a potential difference across the layers, causing the electronic states of the top layer to shift upward relative to those of the down layer \cite{semi-efield1,semi-efield2}. To maintain the same Fermi level across the system, some electrons from the top layer transfer to the down layer. This process accompanies spin flipping, rendering the interlayer compensation incomplete and yielding a non-zero moment.	Applying a $z$-orientated magnetic field has two effects. On the one hand, it forces a portion of spins within each layer to align with the external field direction, as illustrated in the lower-right of Fig. 4. On the other hand, it adds a Zeeman splitting, $\Delta_{\rm H}$, to the already spin-splitting monolayer NbTe$_2$\cite{bfield,Wang-arXiv}. Due to antiferromagnetic interlayer coupling, the sign of $\Delta_{\rm H}$ added to the top and down layers is opposite. This broadens the spin-splitting in the top layer while narrowing it in the down layer. Such asymmetry creates a difference in the Fermi levels between the layers, thereby driving interlayer charge transfer and out-of-plane $P$. Indeed, this interpretation is corroborated by first-principles calculations under applied electric and magnetic fields of 0.25 V/\AA\ and 100 T (see the Supplementary Materials\cite{SI}).

The Fermi-energy-modulation mechanism applies to sliding 1$T$-NbTe$_2$ bilayers. Interlayer sliding creates an effect equivalent to embedding an electric field $E_{in}$ and magnetic field $H_{in}$. When the external field aligns with $E_{in}$/$H_{in}$, it enhances interlayer charge transfer and band spin-splitting, thereby increasing $P$/$M$. Conversely, an external field of opposite direction weakens interlayer charge transfer and band spin-splitting, decreasing $P$/$M$. Likewise, decreasing/increasing $d_{z}$ strengthens/weakens $E_{in}$/$H_{in}$, causing $P$ and $M$ to decrease monotonically with increasing $d_{z}$ [see Fig. 3(c)], but with negligible effect on $\alpha$ [see Fig. 3(d)].

In multiferroic insulators, magnetoelectric coupling manifests as an indirect interaction between spin and charge, typically via spin-charge-lattice or spin-orbit coupling mechanisms\cite{XuML,DongS}. In contrast, within vdW multiferroic metals, e.g., the 1$T$$-$NbTe$_2$ bilayer studied here, magnetoelectric coupling arises from interlayer charge transfer and synchronised spin rearrangement as a result of external field-modulated Fermi levels, not relying on lattice mediation or spin-orbit coupling. It thus represents a direct interaction between spin and charge, which differs fundamentally from the coupling mechanisms involved in multiferroic insulators.

\begin{table}[htbp]
	\centering
	\caption{Summary of ground-state interlayer magnetic coupling (InterM: FM and AFM denote ferromagnetic and antiferromagnetic, respectively; cases in parentheses correspond to non-sliding stacking) and magnetoelectric response for six bilayer multiferroic metals. $\alpha^E_{mod}$ and $\alpha^E_{fir}$ represent the magnetoelectric coupling parameters obtained from Eq. (1) and first-principles linear fitting, respectively. The deviation (Dev) of the former relative to the latter is also provided. In fact, we further compare the effects of different interlayer magnetic configurations on the coupling parameters; these details, along with calculations concerning $\kappa$ and $\eta$, are detailed in the Supplemental Materials\cite{SI}. Noteworthy, Fe$_3$GeTe$_2$ differs from others in that its monolayer does not belong to the 1T phase. For 1T monolayers, only the antiparallel stacked AA' configuration can acquire ferroelectricity through interlayer sliding. However, due to the symmetry of Fe$_3$GeTe$_2$, interlayer sliding produces ferroelectricity in the AA parallel stacking. Thus, we consider AA stacking Fe$_3$GeTe$_2$.}
	\label{table1}
    \renewcommand\arraystretch{1.3}
    \begin{tabular}{c c c c c c c c c c c c c c c}
	\hline
	\hline
		                            &NbTe$_2$  &VSe$_2$    &CrTe$_2$  &MnSe$_2$  &CrS$_2$  &Fe$_3$GeTe$_2$ \\
	\hline
		             InterM         &AFM      &FM(AFM)     &FM        &FM(AFM)    &FM      &FM(AFM)    \\
                   $h$ (\AA)        &10.53     &9.29       &9.63      &8.46      &8.26    &13.27      \\
                   $\kappa$         &1.26     &1.24       &1.28      &1.27      &1.25    &1.19     \\
                   $\eta$           &0.82     &0.088      &0.0091    &0.0033    &0.00049 &0.076   \\
            $\alpha^E_{mod}$          &1.27     &0.15       &0.016     &0.0063    &0.00095 &0.088   \\
            $\alpha^E_{fir}$          &1.41     &0.16       &0.017     &0.0068    &0.0010  &0.095   \\
             Dev (\%)               &10.1     &6.2        &9.0       &7.5       &5.0     &8.1   \\
	\hline
	\end{tabular}
\end{table}

Finally, we elaborate on the generic applicability of the stacking strategy and Fermi-energy-modulation magnetoelectric mechanism. Typically, AA' bilayers exhibit $M_z$ or $M_zT$ symmetry (where $T$ denotes time-reversal symmetry). Despite the antiparallel stacking, the difference in the monolayer magnetization direction permits AA' to exhibit either antiferromagnetic or ferromagnetic interlayer ordering (see Fig. S5\cite{SI}). While a Hamiltonian can be constructed accordingly, its specifics are unnecessary for studying magnetoelectric properties. Without loss of generality, we consider an applied perpendicular electric field. This field breaks $M_z$/$M_zT$ symmetry and induces a net interlayer charge transfer $Q$. By definition, $\alpha^{E} \propto \frac{\partial M}{\partial E} =\frac{\partial M}{\partial Q}\frac{\partial Q}{\partial E}$ (applying the chain rule). Treating AA' as a parallel-plate capacitor yields $\frac{\partial Q}{\partial E} = \epsilon_{0} \kappa h$, where $\epsilon_0$ and $\kappa$ are the vacuum permittivity and material-dependent interlayer permittivity, respectively. Define $\eta$ as the spin polarizability of $Q$, then $\frac{\partial M}{\partial Q}=\frac{2 \eta \mu_B}{e}$, where the factor of 2 arises from spin flipping. Completing the proportionality constant yields
    \begin {equation}
        \alpha^{E} = \frac{2\epsilon_{0}\mu_0\mu_{B}}{e} \frac{\eta \kappa}{h} = C \frac{\eta \kappa}{h},
    \end{equation}
where the constant $C = 2\epsilon_0\mu_0\mu_B/e$. Here $\kappa$ and $h$ are intrinsic attributes of the vdW multiferroics, while $\eta$ is related to $E$ (see the Supplementary Materials\cite{SI}). If $\eta$ varies little within a specific $E$ range, the linear magnetoelectric response dominates.

For further validation, we calculate five additional known metallic magnets, with results summarized in Table I. There are four important findings. First, antiparallel stacking is not a prerequisite for multiferroic metals. Non-1T Fe$_3$GeTe$_2$ with AA parallel stacking also exhibits multiferroic behavior, where the key lies in sliding breaking the inversion symmetry. Second, interlayer magnetic ordering depends not only on the material itself but also on the sliding effect. For VSe$_2$, MnSe$_2$, and Fe$_3$GeTe$_2$, sliding transforms antiferromagnetic ordering into ferromagnetic ordering. Third, $\alpha^E$ values across different multiferroic metals exhibit differences spanning up to three orders of magnitude. This variation primarily originates from the $\eta$ parameter, while changes in $\kappa$ and $h$ are negligible. We find $\eta$ is closely correlated with interlayer magnetic ordering, exhibiting significantly higher values in antiferromagnetic than ferromagnetic ordering (see the Supplementary Materials\cite{SI}). Finally, and most crucially, the Fermi-energy-modulation mechanism we propose is generic, independent of material architecture and interlayer magnetic ordering. The $\alpha^E$ value given by Eq. (1) agrees with first-principles fitting within 5.0\% to 10.6\%. Such precision surpasses that of the (screened) hydrogen model used to calculate exciton binding energies\cite{Olsen,Jiang,WuYJ}, indicating that this mechanism indeed captures the physical essence.
	
To summarize, we propose a general strategy for constructing multiferroic metals. Unlike insulating multiferroics, the magnetoelectric coupling in these multiferroic metals originates from a Fermi-energy-modulation mechanism driven by direct spin-charge interactions. Through first-principles calculations on six known monolayer ferromagnetic metals, we validate the broad applicability of this strategy and mechanism. Although the coupling coefficient falls within the range typical for type-I multiferroics, the purely electronic nature of the magnetoelectric coupling enables these multiferroic metals to exhibit faster response times than lattice-mediated multiferroics, while remaining independent of spin-orbit coupling strength. Moreover, the derived Eq. (1) provides a clear design principle for enhancing response, namely, maximizing the product of $\eta$ and $\kappa$ while minimizing $h$. Our work not only expands material flexibility but also offers a fresh avenue for overcoming the application limitations of insulating multiferroic materials, particularly in fields requiring high-speed response.

	\begin{acknowledgments}
		This work was supported by the Ministry of Science and Technology of China (Grant No. 2023YFA1406400) and the National Natural Science Foundation of China (Grant No. 12474064).
	\end{acknowledgments}

\end{document}